\documentclass[longauth,twocolumn]{aa} 
\usepackage{savesym}
\usepackage{amsmath}
\savesymbol{iint}
\usepackage{txfonts}
\restoresymbol{TXF}{iint}
\usepackage{graphicx}
\usepackage{amssymb}
\usepackage{epsfig} 
\usepackage{times}
\usepackage{natbib}
\usepackage{epstopdf}

\def\apj{ApJ}
\def\apjl{ApJL}

\newcommand\T{\rule{0pt}{2.6ex}}
\newcommand\B{\rule[-1.2ex]{0pt}{0pt}}

\begin{document}

\title{Mrk 421 active state in 2008: the MAGIC view, simultaneous multi--wavelength observations and SSC model constrained}


\author{
 J.~Aleksi\'c\inst{1} \and
 E.~A.~Alvarez\inst{2} \and
 L.~A.~Antonelli\inst{3} \and
 P.~Antoranz\inst{4} \and
 M.~Asensio\inst{2} \and
 M.~Backes\inst{5} \and
 J.~A.~Barrio\inst{2} \and
 D.~Bastieri\inst{6} \and
 J.~Becerra Gonz\'alez\inst{7,}\inst{8} \and
 W.~Bednarek\inst{9}
% ...
}
\institute { IFAE, Edifici Cn., Campus UAB, E-08193 Bellaterra, Spain
 \and Universidad Complutense, E-28040 Madrid, Spain
 \and INAF National Institute for Astrophysics, I-00136 Rome, Italy
 \and Universit\`a  di Siena, and INFN Pisa, I-53100 Siena, Italy
 \and Technische Universit\"at Dortmund, D-44221 Dortmund, Germany
 \and Universit\`a di Padova and INFN, I-35131 Padova, Italy
 \and Inst. de Astrof\'{\i}sica de Canarias, E-38200 La Laguna, Tenerife, Spain
 \and Depto. de Astrof\'{\i}sica, Universidad de La Laguna, E-38206 La Laguna, Spain
 \and University of \L\'od\'z, PL-90236 Lodz, Poland
% ...
}

\date{Received  08 June 2011; accepted 07 January 2012}

\abstract{The blazar Markarian 421 is one of the brightest TeV gamma--ray
sources of the northern sky. From December 2007 until June 2008
  it was intensively observed in the very high energy (VHE, $E > 100\, {\rm GeV}$)  band by the single--dish
  Major Atmospheric Gamma--ray Imaging Cherenkov telescope (MAGIC-I).} 
{We aimed to measure the physical parameters of the emitting region of
    the blazar jet during active states.}
{We performed a dense monitoring of the source in VHE with MAGIC--I, and also
  collected  complementary data in soft X--rays and optical--UV bands; then, we
modeled the  spectral energy distributions (SED) derived from simultaneous multi--wavelenght data within the synchrotron
self--compton (SSC) framework.} 
{The source showed intense and prolonged
  $\gamma$--ray 
  activity during the whole period, with integral fluxes ($E > 200$ GeV) seldom below the
 level of the Crab Nebula, and up to 3.6 times this value. Eight datasets of simultaneous optical--UV (KVA, {\it Swift}/UVOT), soft X--ray
({\it Swift}/XRT) and MAGIC-I VHE data were obtained during different outburst
phases. The data constrain the  physical parameters of the jet, once the
spectral energy distributions obtained are interpreted within the framework of a  single-zone SSC leptonic model.}
{The main outcome of the study is that within the homogeneous model high
Doppler factors ($40 \leq \delta \leq 80$) are needed to reproduce the
observed  SED;  but this model cannot explain the observed short time--scale
variability,  while it can be argued that inhomogeneous
models could allow for less extreme Doppler factors, more intense magnetic fields
and shorter electron cooling times compatible with hour or sub-hour scale variability.}


\titlerunning{Mrk 421 active state in 2008 with MAGIC}
\authorrunning{Aleksi\'c et al.}
                            
\maketitle


\section{Introduction}

Blazars, a common term used for flat--spectrum radio quasars (FSRQ) and
BL Lacertae objects, constitute the subclass of active galactic nuclei
  (AGN) that
  is most commonly detected  in the Very High Energy (VHE, $E >  100\, {\rm GeV}$) $\gamma$--ray
    band. In these  sources the dominant radiation component
originates in a relativistic jet pointed nearly toward the
observer.  The double-peaked spectral energy distribution (SED) of blazars is
attributed to a population of relativistic electrons spiraling in the
magnetic field of the jet. The low--energy peak is commonly thought to be causedby
synchrotron emission, because of its spectrum and polarization.
The second, high energy peak is attributed to inverse
Compton scattering of low--energy photons in leptonic acceleration models \citep{1992ApJ...397L...5M,dermer, bloom}. 
Alternative models invoking a relevant contribution from accelerated hadrons can also
sufficiently describe the observed SEDs and light curves
(\citealt{1993A&A...269...67M,Mucke} but see \citealt{2009ApJ...704...38S}
on FSRQs). 

Blazars are highly variable in all wavebands and the relation
between variability in different bands is a key element in
distinguishing between different models. For instance, homogeneous leptonic
models predict correlated variability between, e.g., X rays and $\gamma$ rays,
which is already observed in  high--frequency peaked BL Lacs (HBL; see e.g. \citealt{2008ApJ...677..906F} on
Mrk 421 itself). 
On the other hand, phenomena such as the "orphan" flare from 1ES1959+650
reported in \cite{2004ApJ...601..151K} or the ultrafast ($\sim$ hundreds of seconds) events occasionally observed in
some sources \citep[e.g.][]{2007ApJ...664L..71A,2007ApJ...669..862A} are harder to explain  whithin this frame. 

 Among blazars, HBLs are the most often observed subsample in the VHE domain,
 because the high--energy bump peaks  at GeV--TeV energies, while the low--energy peak is located at UV to X--rays energies \citep{padovani07}. 
 This makes HBLs, such as Mrk 421, ideal targets for  sensitive, low--energy threshold imaging air
Cherenkov telescopes (IACT) such as MAGIC--I, in combination with
soft X--ray telescopes, which observe the synchrotron bump instead;
this combination of instruments samples the source SED and unravel the regions
of the two peaks, the most valuable tracers of the source state. 
\\
\\
Mrk 421  is one of the closest ($z =0.031$, \citealt{1991trcb.book.....D}) and
brightest extragalactic TeV sources; therefore it was the first to be detected
\citep{1992Natur.358..477P} and remains one of the best studied. 
The VHE integral flux can vary from a few tenths to a few Crab Units (e.g., see
\citealt{2009ApJ...691L..13D,2009arXiv0907.0893H} or \citealt{2009arXiv0908.0010P}), on time
scales as short as 15 minutes \citep{1996Natur.383..319G}. 
The $\nu F(\nu)$
distribution of the emitted photons follows the standard "double--bumped"
shape, but varies significantly from low--activity states to the most intense
flares, on time scales that in X-rays can be  of few hours \citep{2009ApJ...699.1964U}.
The low-energy bump peaks in the $0.1-10$ keV range (see
e.g. \citealt{2008ApJ...677..906F}), as is usual for HBLs; the
maximum of the high--energy bump is usually found below 100 GeV, but can also
 move around according to the state of the source,  usually following a
``harder-when-brighter" behavior \citep{Krennrich02,2008ApJ...677..906F,Acciari11,2007ApJ...663..125A}
analogous to that traced by the X-ray emission \citep[e.g.][]{Brinkmann05,Tramacere07,Tramacere09}.

This peculiar SED shape favors multi--wavelength (MWL) studies that exploit the MAGIC--I sensitivity and low--energy threshold in VHE, and
soft X--ray telescopes. MAGIC--I can detect Mrk 421 at the $5 \sigma$ level with
exposures as short as a few minutes, depending on the source brightness.
 In the soft X--ray domain  the All--Sky Monitor (ASM) onboard the
{\it Rossi X--ray Timing Explorer} (RXTE) can continuously provide a
daily averaged flux, while the {\it X--Ray Telescope} (XRT) onboard the {\it Swift}
satellite can observe the source with far better precision and
energy resolution in $\sim1$ ks targeted exposures.
From the observation of the source spectrum in both X--ray and VHE  a unique set of physical
parameters that describe the source can be derived within a single-zone
synchrotron self--Compton (hereafter, SSC) model, following
\cite{1998ApJ...509..608T} . An SSC modeling of the SED of Mrk 421 has been already performed in the
past (see e.g. \citealt{1997MNRAS.292..646B};
\citealt{1998ApJ...509..608T}; \citealt{1999ApJ...526L..81M} or the more recent
\citealt{2008ApJ...677..906F}). 

Lately automated $\chi^2$ minimization
procedures \citep{2008ApJ...686..181F,2011ApJ...733...14M} have been applied.

The main limitations to previous works came from
the use of the  former IACTs such as Whipple
\citep[e.g. ][]{2008ApJ...677..906F} or HEGRA
\citep[e.g. ][]{2000ApJ...542L.105T}, which were characterized by a higher energy
threshold and worse sensitivity at VHE. This in turn led to poor sampling of
the IC peak region, basically limited to the less informative, steeply decaying
hard energy tail of the bump; moreover, integration over different nights of
observation was commonly needed to obtain a significant VHE spectrum, thus
yaveraging out the SED evolution,  such as in \citet{1992ApJ...397L...5M}. 
 In 2008 Mrk 421 went through a long and intense outburst phase,
characterized by VHE fluxes quite constantly above the Crab level and
superimposed shorter and brighter flares; a remarkably dense
follow--up of this evolution was possible in optical, X--rays and VHE with MAGIC-I.
The main outcome of this campaign is that eight tightly contemporary
observations of Mrk 421 in optical, X--rays and VHE $\gamma$--rays of active
states could be achieved. This allowed the reconstruction and modeling of the
optical-UV/X-ray/TeV MWL SED of Mrk 421 on short time scales ($\sim 1$
  hour), the main improvement in this work with respect to the past literature on
  the subject.
 A similar approach was followed, for the same Mrk 421, by the
VERITAS Collaboration \citep{Acciari11}.

The paper is structured as follows: in Section \ref{observations} we report on
the observations and data analysis; 
 in Section  \ref{results} we report the VHE light curve and spectra,  and complementary results in X rays and optical--UV band; in Section \ref{simultaneous} we build the  SED of Mrk 421 in the eight states for which a set
of simultaneous MWL data was available, which we model in the
framework of a standard one--zone SSC model; finally we discuss the results in Section \ref{discussion}.

\section{Observations and data analysis}
\label{observations}
The timespan of the MAGIC VHE observations of Mrk 421 reported here began in
December 2007 and ended in June 2008. Contemporary data from other instruments
are considered for the multi--wavelength analysis, namely soft X--ray data from
RXTE/ASM, and {\it Swift}/XRT, optical--UV data from {\it Swift}/UVOT and optical
R--band data from the Tuorla Observatory. A summary description of the
instruments, the datasets and the analysis follows. 

\subsection{MAGIC-I VHE observations}

 MAGIC\footnote{Major Atmospheric Gamma--ray Imaging Telescope}-I  (formerly
MAGIC) is an IACT located on
the western Canarian island of La Palma, at the Observatory of Roque de Los
Muchachos ($28.75^\circ$ N, $17.89^\circ$ W, $2225\ $m a.s.l.). With its
tessellated parabolic mirror ($D= 17\,$m, $f/D=1$), it has been the largest
single--dish IACT in operation from late 2004 until the advent in 2009 of MAGIC--II, a
twin (but substantially improved in many respects) telescope; since then the two
telescopes are operated as
a stereo IACT system (MAGIC Stereo). Its 234 m$^2$  surface allowed for the lowest energy threshold
among IACT systems at that time: the trigger threshold of the telescope at the epoch of
this campaign reached as low as $60\,$ GeV for observations close to the
zenith in optimal conditions.
 A detailed description of the telescope and  data analysis can be found in
dedicated papers (e.g. \citealt{2004NIMPA.518..188B}; \citealt{2005ICRC....5..359C}; \citealt{crab}).
All MAGIC--I observations  considered for the present study were performed
following a major hardware upgrade \citep{2008ICRC....3.1481G} that was completed in February
2007, which enhanced the time sampling capability of
the data acquisition (DAQ) from 300 MHz to 2 GHz. This allowed a better rejection of the night sky background (NSB)
and introducing new refined analysis techniques
\citep{2008ICRC....3.1393T} based on the time properties of Cherenkov
signals,  which lowered the integral sensitivity for point sources down to
  1.6 \% of the Crab Nebula flux \citep[for a 5$\sigma$ significant detection in
  50 hours, above 280 GeV;][]{timing}. 

During the observation period considered here, MAGIC observed Mrk 421 for a total
of 81 nights, with exposure times ranging 
from $\sim$ 20 to $\sim$ 240 minutes. 
 This comprised both short untriggered observations, aimed to an unbiased sampling
of the source state (studied in detail in \citealt{blazarmonitoring}), and
deeper extended observations of peculiar states, triggered either by the
former or by external alerts from other bands.
All observations were performed in the false--source tracking
(``wobble'', \citealt{1994APh.....2..137F}) mode. The
method consists of alternatively tracking two positions in the sky that are
symmetrical with respect to the
source nominal position and 0.4$^\circ$ away from it.

The MAGIC--I data were analyzed using the standard analysis chain described in 
\cite{crab,NIMA} and \cite{timing}. 
Preliminary quality checks were performed to exclude poor quality
data, such as those owing to bad weather or occasional technical
problems. Furthermore, the dataset was restricted to observations performed  under
 dark conditions, and in the range of zenith angle ranging from $\sim$
 5$^\circ$ at culmination to 46$^\circ$. 
A cleaning algorithm involving the time structure of the shower images was then
applied, which further selected the events and removed the NSB contribution to the images. Surviving images were parametrized in terms
of the extended set of Hillas parameters \citep{1985ICRC....3..445H} described in the mentioned literature.   
To suppress the
unwanted background showers produced by charged
cosmic rays, a multivariate classification
method known as random forest (RF, \citealt{breiman}) was implemented and  applied
\citep{randomforest}. 
 An analogous procedure allowed the estimation of
the energy of the primary $\gamma$--rays.
The signal extraction was performed by applying cuts in the \textsc{Size},
\textsc{Hadronness} and \textsc{Alpha} parameters described in the aforementioned literature. In particular the
\textsc{Size} cut,  which we set to reject events with less than 150 photoelectrons of total
charge, implied  an energy threshold $\sim 140 \ $  GeV in the present analysis.
A total excess of $\sim 48 \times 10^3$  events from the selected $\sim 60$ hours of
observation was detected. 
 The whole analysis procedure was validated step by step on compatible
 datasets from observations of the Crab Nebula. 

\subsection{Optical, UV and X--ray observations}
The {\it Swift} satellite \citep{2004ApJ...611.1005G} is a NASA mission,
launched in 2003, devoted to observations of fast transients, namely
prompt and afterglow emission of gamma--ray bursts. These are detected with the monitoring coded mask Burst
Alert Telescope (BAT, \citealt{2005SSRv..120..143B}) which is sensitive
to 15-150 keV X--rays and covers a wide field of view (FoV) with a resolution of few
arcminutes, and then rapidly targeted  with the two co--aligned  pointing
instruments, X--Ray Telescope (XRT, \citealt{2005SSRv..120..165B}) and
Ultra--Violet Optical Telescope (UVOT, \citealt{2005SSRv..120...95R}).

 The fast repositioning capability of the spacecraft allows snapshots of
variable sources with little overheads. For Mrk 421,
observations lasting $\sim 1$ ks allow the derivation of a detailed X--ray
spectrum and multi--filter optical--UV photometry because of the sensitivity of
the targeted instruments and the brightness of the source.

{\it Swift}/XRT is a Wolter type--I grazing incidence telescope, with  110 cm$^2$ effective
area, 23.6'  FoV and 15'' angular resolution, sensitive in the 0.2--10 keV
energy band.
During the MAGIC campaign the instrument performed 43 targeted X--ray
observations of Mrk 421 of  typical exposure times 1--2 ks.
{\it Swift}--XRT data were
reduced using the software distributed with the {\tt  heasoft} 6.3.2 package  by the NASA High Energy Astrophysics Archive
Research Center (HEASARC). The {\tt xrtpipeline}  was set for the photon
counting or window timing modes  and single pixel events (grade $0$) were selected.

UVOT is a 30 cm diffraction--limited  optical--UV telescope, equipped with
six different filters, sensitive in the 1700--6500 \AA\  wavelength range, in a
17' $\times$ 17' FoV. Unfortunately, during the January 2008 campaign UVOT
did not observe the source, so that the UVOT datasets were fewer than the XRT
pointings and therefire five out of the eight datasets studied in Section \ref{simultaneous}) have no contemporary UVOT 
observations.
Therefore we restricted the analysis of UVOT data to the three observations
simultaneous with MAGIC-I performed on February 11 and April 2 and
3 with the UV filters alone.
 The analysis was performed with  the \texttt{uvotimsum} and \texttt{uvotsource} tasks with a source region of $5''$, while the background was extracted from a source--free circular region with radius equal to $50''$ (it was not possible to use an annular region because of a nearby source).  
The extracted 
 magnitudes were corrected for Galactic
extinction  using the values of \citet{1998ApJ...500..525S} and applying the
formulae by \citet{1992ApJ...395..130P} for the UV filters, and eventually
were converted into fluxes following \citet{2008MNRAS.383..627P}.

The All-Sky Monitor (ASM) onboard the {\it Rossi X-ray Timing Explorer} (RXTE, \citealt{1993A&AS...97..355B}) is
sensitive enough to set one point per day from Mrk 421, which means a poorer
precision, but denser coverage than {\it Swift}/XRT; 
 
 The publicly available ASM data products were taken from the results provided by the ASM/RXTE teams at MIT and at the RXTE SOF and GOF at NASA's GSFC.

The Tuorla Observatory constantly monitors the MAGIC VHE (known or potential)
target sources, with the 35\ cm remotely operated Kungliga Vetenskaplika
Academy (KVA) optical telescope that is also located at Roque de los Muchachos and with a $103\,$ cm telescope located
at Tuorla, Finland.  During the period of the MAGIC--I observations, 117
photometric measurements of Mrk 421 were obtained in the Johnson $R$--band.
The optical data were reduced by the Tuorla Observatory as described in \cite{2007A&A...475..199N}.
 The light contribution from the host galaxy and nearby companion galaxy
 ($F_{h+cg} =8.07\pm 0.47\,$ mJy) was subtracted from the measured
 fluxes.

\section{Results}
\label{results}
\subsection{MAGIC-I VHE  light curves}
\begin{figure*}[!th]
  \centering
\includegraphics[width=\textwidth]{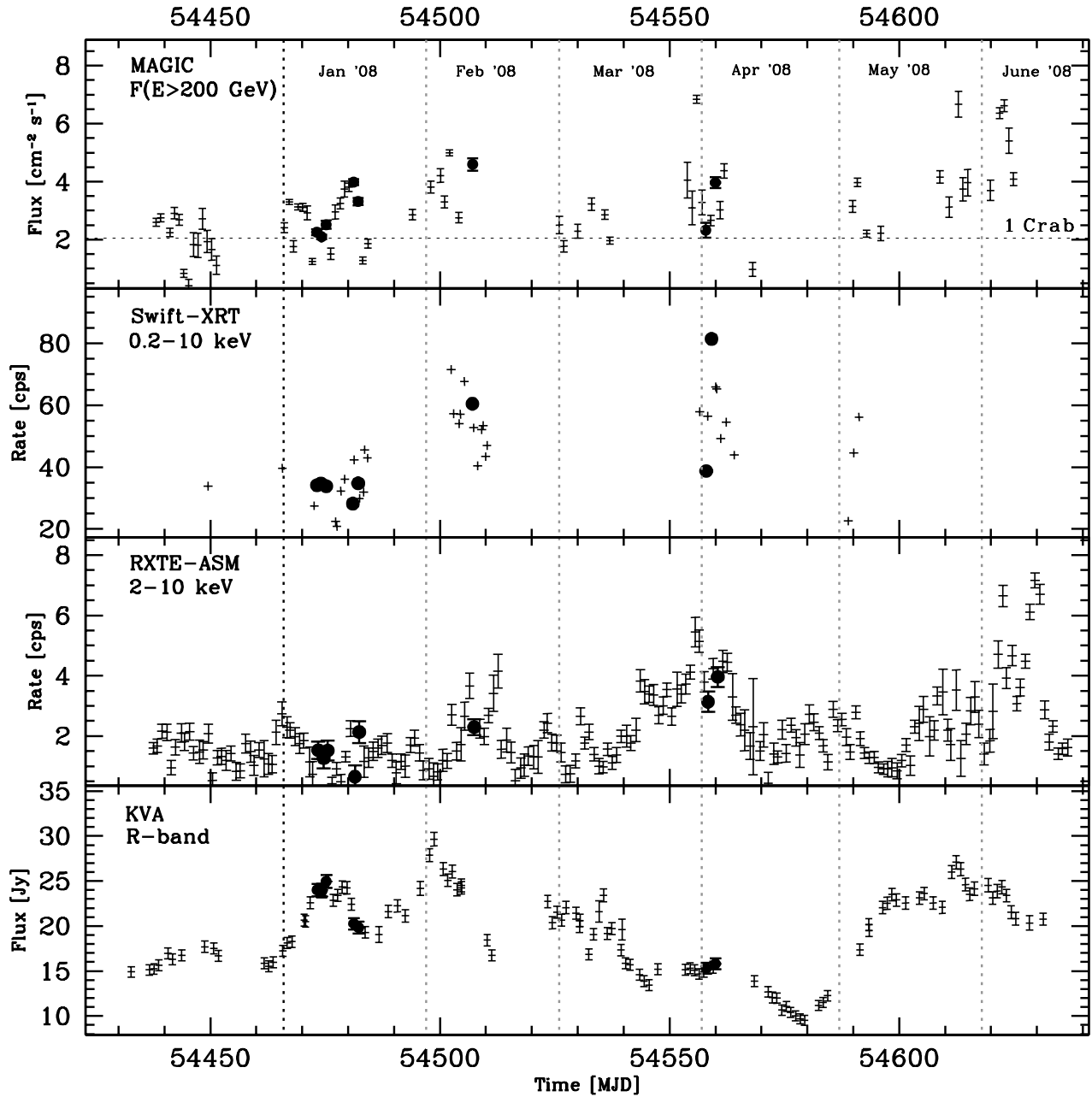}
  \caption{Multi--wavelength light curves of Mrk 421 along the MAGIC--I observation
    period; full circles mark the fluxes observed when MAGIC--I  and {\it Swift}/XRT
    were pointed at the source simultaneously. Upper panel: MAGIC--I VHE light curve
    above 200 GeV, for the 66 observation nights that passed quality
    cuts. MAGIC--I detected the source clearly in all nights; the integral
    flux was below the Crab Unit (C.U., $\approx F_{E>200\,GeV} =
    2.0 \times 10^{-10}$ photons $\cdot cm^{-2}s^{-1}$, represented here by the
    dashed horizontal line) only in a few nights. A maximum flux of $\sim 3.6$
    C.U. was observed on  2008 March the 30th
    (MJD=54555). Middle--upper panel: soft X--ray (0.2--10 keV) count rates measured by {\it Swift}/XRT. Middle--lower
    panel: soft  X--ray (2--10 keV) count rates measured by RXTE/ASM. Lower
    panel: Johnson R--band optical light curve from the Tuorla Observatory.}
  \label{MWL_LC}
 \end{figure*}

The night-averaged integral flux above a conservative threshold of 200 GeV was
calculated for each of the 66 nights with datasets that survived the quality
cuts. The VHE light curve of
Mrk 421 along the campaign is plotted in the top panel of Figure \ref{MWL_LC}.
Interestingly, although Mrk 421 is believed to emit a low VHE flux
baseline \citep{1996ApJ...460..644S}, the flux was seldom below one Crab Unit
 (hereafter C.U. corresponding to an integral flux $F_{E>200\, {\rm GeV}} = 1.96\pm 0.05_{stat} \times 10^{-10}
{\rm cm}^{-2} {\rm s}^{-1}\ $, \citealt{crab}) for the entire period,
confirming an intense and persistent active state. The maximum observed flux
($F_{E>200\, {\rm GeV}} = 6.99\pm 0.15_{stat} \times 10^{-10} {\rm cm}^{-2}
{\rm s}^{-1}$) was on  2008 March 30.  Similar fluxes were reached in
  the flare that occurred in June 2008, which was already studied in detail in
  \citet{2009ApJ...691L..13D}. 
 Similar high flux levels during the same period are reported in
  \citet{Acciari11}, with a record flux of $\sim 12 \times 10^{-10} {\rm
    cm}^{-2}{\rm s}^{-1}$ above 300 GeV (corresponding to $\sim$ 10 C.U.) observed in May 2008.

\begin{figure*}
  \centering
  \includegraphics[width=5.71in]{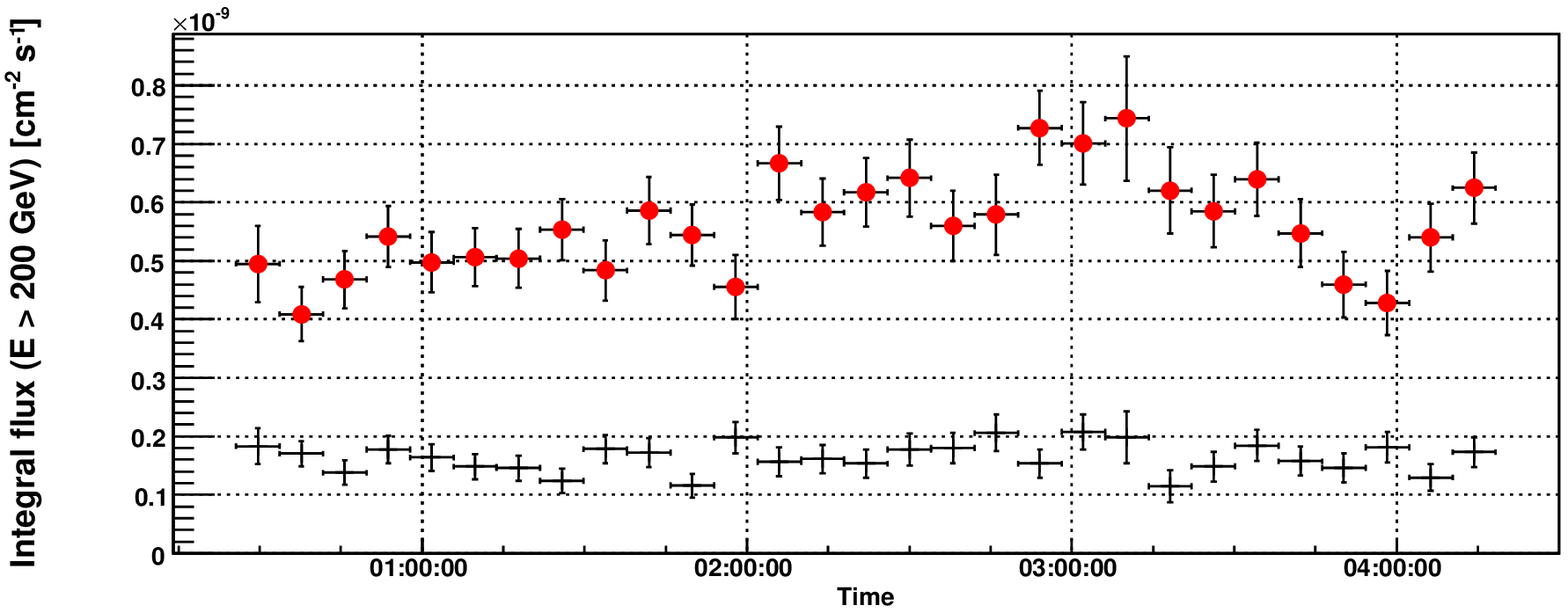}
\includegraphics[width=5.71in]{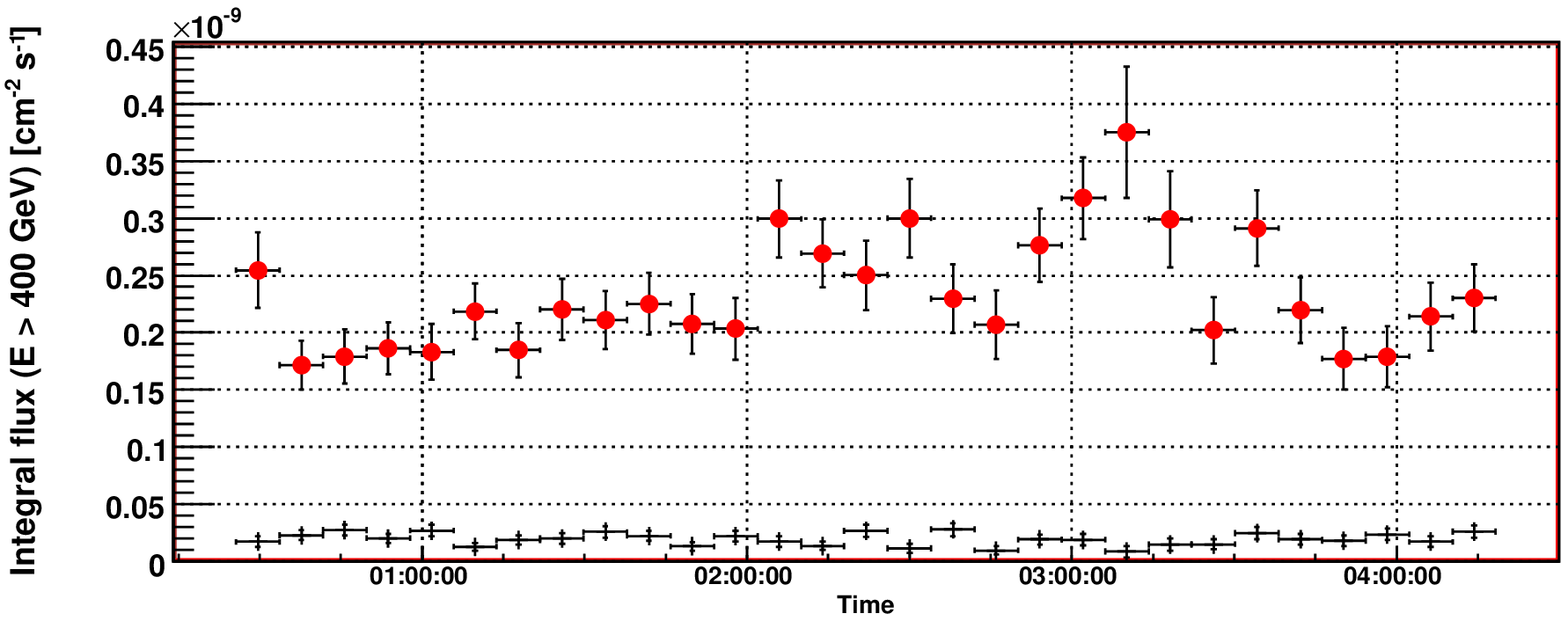}
  \caption{Mrk 421 VHE light curves in 8--minute time bins from the observations taken on 2008 February 6.  Integral flux of excess (filled circles) and background (thin crosses) events are plotted. The energy threshold is 200 GeV (upper panel) and 400 GeV (lower panel).}
\label{LC_20080206}
\end{figure*}

 The sensitivity of MAGIC--I allowed us to investigate the sub--hour scale
  evolution of VHE flux for  Mrk 421 in high state, searching for the rapid variations already
  reported in literature \citep{1996Natur.383..319G}. The most
  interesting result was obtained  on 2008 February 6, when a  long ($\sim$ 4
  hours) observation of a high ($\sim 2.5$ C.U. above 200 GeV) state was  performed.  
The VHE  light curve in 8--minute time bins is shown  in Figure \ref{LC_20080206},
above a softer ($E > 200\, {\rm GeV}$, upper panel) and harder ($E> 400\, {\rm
  GeV}$, lower panel) energy threshold. An episode of variability with
doubling/halving times down to 16 minutes can be seen with the harder cut. The
hypothesis of a steady flux is unfit in both light curves according to
results of a $\chi^2$ test, giving $\chi^2/ {\rm d.o.f.}$ of $55/28$ (probability below 0.2\%) and $63/28$
(probability $\sim 10^{-4}$), respectively, which confirms
variability on the scale of hours or less.
Unfortunately, no simultaneous Swift/XRT observation was performed in this
night,  therefore  we
could not add this VHE observation  to the set of simultaneous MWL SED.
 No firm conclusion could be drawn on the sub--hour variability for the
 simultaneous datasets, because some observation
windows were very short (e.g. April the 2nd and 3rd) and other were made in lower flux levels
(e.g. January 8, 9, 10), which led to poorer event statistics.

\subsubsection{Multi--wavelength data}

 The {\it Swift}-XRT observed count rates in the $0.2-10\,$  keV band are
 reported in the middle--upper panel of Figure \ref{MWL_LC}.

The count rates observed by RXTE/ASM in the $2-10\,$ keV band are shown in the middle--lower panel of Figure \ref{MWL_LC}.

The R--band optical light curve from KVA observations is reported in the bottom panel of Figure \ref{MWL_LC},
while the available measurements related to the simultaneous datasets listed
in Table \ref{421_tight_tab} (see Section \ref{simultaneous}) are plotted in
Figure \ref{seds} after correction for Galactic extinction, again applied
according to the values of \citet{1998ApJ...500..525S}.

\subsection{ VHE spectra derived from MAGIC-I data}

We restricted the study of the spectra to the subset of the eight observations
of interest for modeling the MWL SED (see Section \ref{simultaneous}), listed in Table \ref{421_tight_tab}.
From each observation we derived a VHE spectrum in bins of the estimated
energy of the $\gamma$--ray primary events.
Then we applied the Tikhonov unfolding algorithm \citep{2007NIMPA.583..494A} to reconstruct the physical spectrum in terms of the true
energy of the primary $\gamma$--rays. 
A best fit to the data was then performed, assuming a log--parabolic model for
the differential spectrum:
\begin{equation}
\frac{dN}{dE\, dA\, dt}= f_0 \times (\frac{E}{E_0})^{(a + b \cdot \log{(\frac{E}{E_0}}))} ,
\end{equation}
where 
the pivot energy $E_0$ is chosen 300 GeV in the present case. In
  three cases a simple power law was sufficient to fit the data.

 For each night, the integral VHE flux above 200 GeV, the parameters of
the fit  to the observed (no EBL correction) emission and the
$\tilde{\chi }^2$ are reported in Table \ref{421_tight_result}; quoted uncertainties are statistical only.

\begin{figure}[!ht]
\centering
\includegraphics[width=3.in]{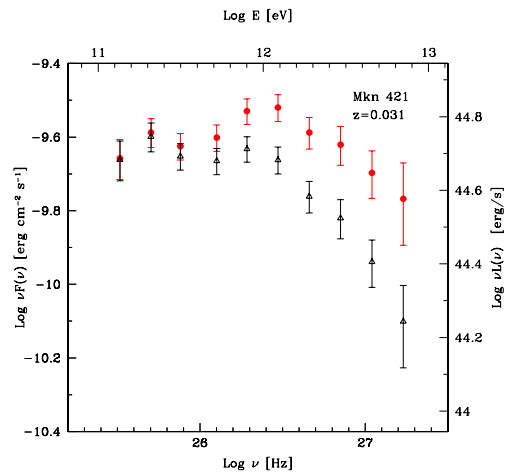}
  \caption{VHE SED of Mrk 421 derived from the MAGIC-I observations performed on
    2008 March 30,
    when the flux rose to 3.6 Crab Units.  Data points before (black open triangles) and after
    (red filled circles) applying a correction for EBL absorption following
    \citealt{2008A&A...487..837F} are shown. The observed position of the IC peak
    is evaluated at $\sim 500$ GeV from the fit with a curved power law, and
    above 1 TeV after deabsorption. This VHE spectrum was the hardest among
    the ones studied here, as illustrated in Figure \ref{VHE_spectra_MAGIC}.}
\label{421_0331}
\end{figure}

\begin{figure*}[!ht]
  \centering
\includegraphics[width=0.9\textwidth]{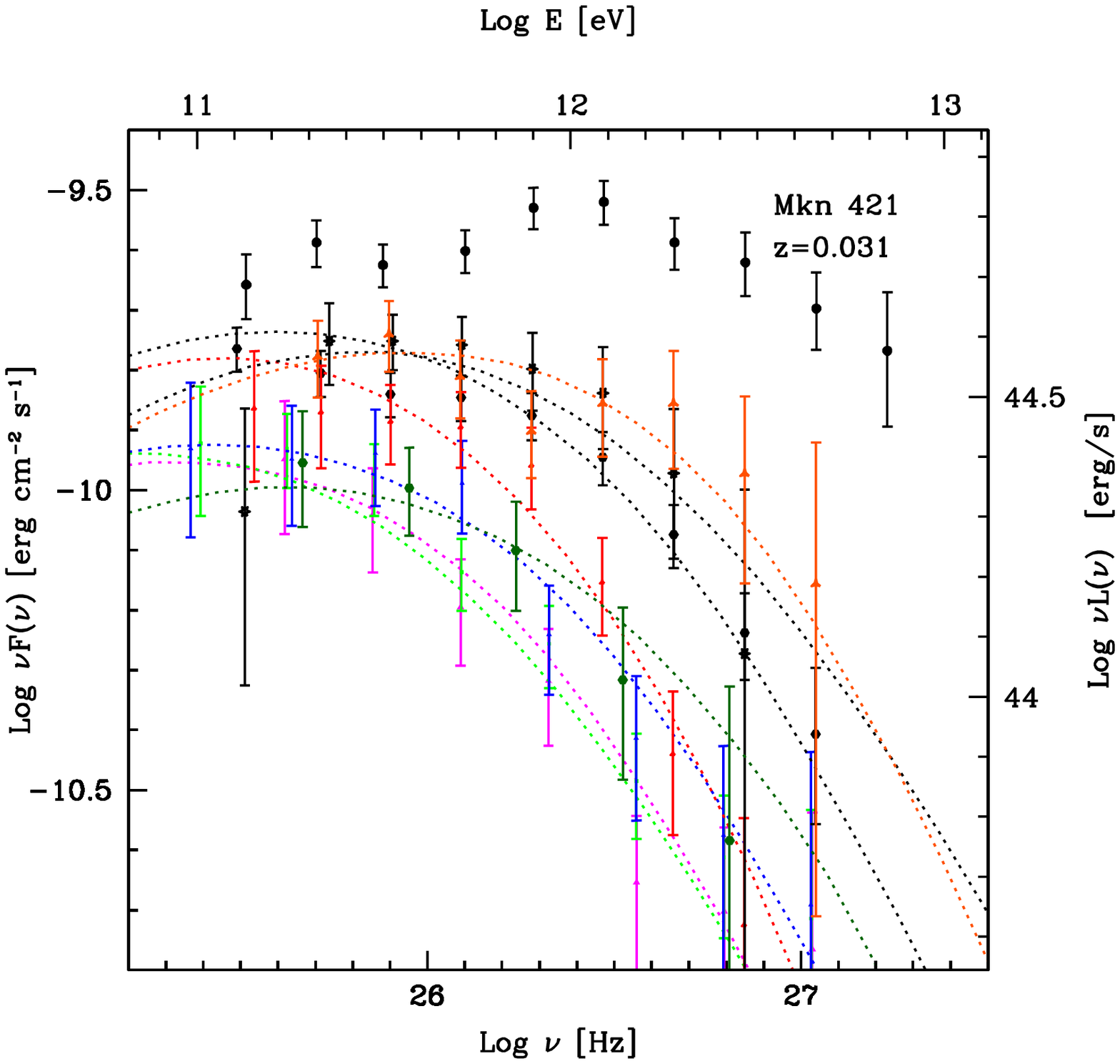} 

  \caption{VHE SED of Mrk 421 derived from the MAGIC--I observations performed
    in the eight time slots with tightly simultaneous Swift/XRT data (see Section \ref{simultaneous}). The spectra    are  shown after correction of the EBL absorption, following
    \citet{2008A&A...487..837F}. For comparison, the spectrum derived from
    the observation that registered the highest flux (3.6 C.U. above 200 GeV)
    of the whole campaign, performed on 2008 March 30. For each night we also
   plot our model (see Section \ref{sedmodeling}) to guide the eye.}
\label{VHE_spectra_MAGIC}
\end{figure*}

 Adopting the $a$ parameter, giving the slope of the spectrum at the pivot
  energy, as an estimator of its hardness, it is evident that the well--observed "harder when
  brighter'' trend \citep[see e.g.][]{Acciari11,2008ApJ...677..906F} is nicely reproduced in these spectra.
The spectral points were subsequently corrected for extra--balactic background
light (EBL) absorption.  The \citet{2008A&A...487..837F} EBL model has been assumed
 in this work, 
even if consistent results can be obtained with other models such as the
more recent \citet{2011MNRAS.410.2556D} model, given that  for this close--by
source the model-to-model differences in opacity below 10 TeV  are
 dominated by the statistical uncertainties in hour-scale integrated VHE
spectra.
 The data points are plotted in Figure \ref{VHE_spectra_MAGIC} along with the SSC models (see
Section \ref{sedmodeling}) that are anticipated here only as a help to guide
the eye. 
 For comparison we also plot, without a model, the SED built from the
observation achieving the highest VHE flux of the whole campaign (March
30). Unfortunately, we were unable to include this interesting dataset in the SED study,
because {\it Swift} could only observe with 14 hours of delay with respect to MAGIC--I.
 Anyway, the VHE spectrum derived from this observation is intriguingly
hard, peaking around 500 GeV, well within the MAGIC-I band.  In
Figure \ref{421_0331} the observed SED (black open triangles)  is plotted together
with the deabsorbed one (red filled circles), which peaks above 1 TeV. 

\begin{table*}[th!]
\centering
\begin{tabular}{lccccc}
\hline
\hline

Night \T  \B &Integral flux &$f_0$&$a$&$b$&  $\tilde{\chi}^2/$d.o.f.\\                                       
& cm$^{-2}$s$^{-1}$&cm$^{-2}$s$^{-1}$ TeV$^{-1}$&&&\\                                     
yyyy-mm-dd&[$\times 10^{-10}$]&[$\times 10^{-10}$]&&&\\                                   
&($E>200$ GeV)&($E_0=300$ GeV)&&&\\                                                   
\hline                                                                               
 2008-01-08 \T &$2.13\pm 0.20$&$5.9\pm0.7         $&$-2.72\pm0.12$&$      -     $&$0.40/5$\\
 2008-01-09&$2.61\pm 0.11$&$6.3\pm0.3         $&$-2.50\pm0.07$&$-0.44\pm0.15$&$0.63/7$\\
 2008-01-10&$2.53\pm 0.16$&$7.4\pm0.5         $&$-2.42\pm0.08$&$-0.52\pm0.20$&$1.26/6$\\
 2008-01-16&$4.42\pm 0.14$&$ 10\pm1\phantom{0}$&$-2.25\pm0.07$&$-0.33\pm0.10$&$0.34/6$\\
 2008-01-17&$3.80\pm 0.19$&$9.8\pm1.2         $&$-2.37\pm0.10$&$-0.57\pm0.18$&$0.70/6$\\
 2008-02-11&$5.34\pm 0.32$&$ 12\pm1\phantom{0}$&$-2.11\pm0.14$&$-0.44\pm0.24$&$1.10/6$\\
 2008-04-02&$2.94\pm 0.32$&$7.1\pm0.5         $&$-2.44\pm0.16$&$    -       $&$0.31/3$\\
 2008-04-03 \B &$4.53\pm 0.30$&$ 11\pm1\phantom{0}$&$-2.35\pm0.10 $&$    -      $&$0.37/6$\\
\hline
\hline
\end{tabular}
\caption{Results from MAGIC--I VHE observations of Mrk 421 during the eight nights with
  simultaneous \textit{Swift}/XRT data.}
\label{421_tight_result}
\end{table*}

\subsection{Soft X--ray spectra derived from {\it Swift}/XRT data}

For the eight simultaneous observations with MAGIC--I listed in Table
\ref{421_tight_tab} we extracted the {\it Swift}/XRT spectra  to build the MWL SED (see
Section \ref{simultaneous}). Data were rebinned to obtain at least 
30 counts per energy bin. 
Broken power--law models were fitted to the spectra in the range 0.35-10 keV. 
The X--ray reddening due to absorbing systems along the light
travel path was corrected assuming the Galactic value for the column density
of neutral hydrogen $N_{H} =1.6 \times 10^{20}$ cm$^{-2}$
\citep{1995ApJS...97....1L}.
 Table \ref{421_swift_fit} reports for each dataset (with
  uncertainties in parentheses) the Obs ID, the UTC time at the beginning of observation, the exposure time,
  the integral flux in the 2--10 keV band, the spectral indexes, break energy and
  normalization at 1 keV of the broken power law, and the resulting
  $\tilde{\chi}^2$ (reduced $\chi^2$) with the number of degrees of freedom.

\begin{table*}[th!]
\begin{tabular}{ccccccccc}
\hline
\hline
Obs. ID \T &Start&Obs.&$F_{2-10 {\rm keV}} $&$\alpha_1$&$
E_{break}$&$\alpha_2$& $f_0$& $\tilde{\chi}^2$/{\rm d.o.f.}\\
&Time (UT)&Time&erg$/$cm$^2/$s&&keV&&&\\
&yyyy-mm-dd hh:mm \B &ks& [$\cdot 10^{-12}$] &&&&&\\
\hline
\hline
00030352041 \T &2008-01-08 02:30& 2.0 & 280 & 2.31(0.03) & 1.20(0.09) &2.58(0.03) & 0.242(0.003) & 1.34/163 \\
00030352042&2008-01-09 04:04& 2.0 & 283 & 2.28(0.03) & 1.05(0.10) & 2.60(0.03) & 0.257(0.005) & 1.30/170 \\
00030352044&2008-01-10 02:27& 2.3 & 284 & 2.32(0.02) & 1.10(0.10) & 2.57(0.02) & 0.245(0.003) & 1.59/178 \\
00030352053&2008-01-16 03:21& 1.2 & 345 & 2.19(0.04) & 1.24(0.18) & 2.45(0.04) & 0.242(0.004) & 1.16/122 \\
00030352055&2008-01-17 03:29& 0.8 & 311 & 2.21(0.03) & 1.97(-0.18/+0.4) & 2.75(-0.09/+0.19) & 0.243(0.003) & 1.33/101 \\
00030352068&2008-02-11 03:40& 1.9 & 587 & 2.19(0.01) & 2.43(0.2) & 2.57(0.06) & 0.372(0.002) & 1.76/215 \\
00030352083&2008-04-02 00:42& 0.9 & 474 & 2.09(0.02) & 2.86(0.33) & 2.51(-0.08/+0.15) & 0.260(0.003) & 1.47/130 \\
00030352086 \B &2008-04-03 21:59& 1.2 & 961 & 1.95(0.02) & 2.37(-0.16/+0.28) & 2.33(0.06) & 0.438(0.003) & 1.48/210 \\
\hline
\hline
\end{tabular}
\caption{Mrk 421 soft X--ray fluxes and spectral parameters from the eight \textit{Swift}/XRT
  datasets simultaneous to  MAGIC--I observations.}
\label{421_swift_fit}
\end{table*}

\section{Simultaneous multi--wavelength datasets}
\label{simultaneous}
\begin{figure*}
  \centering
\includegraphics[width=0.75\textwidth]{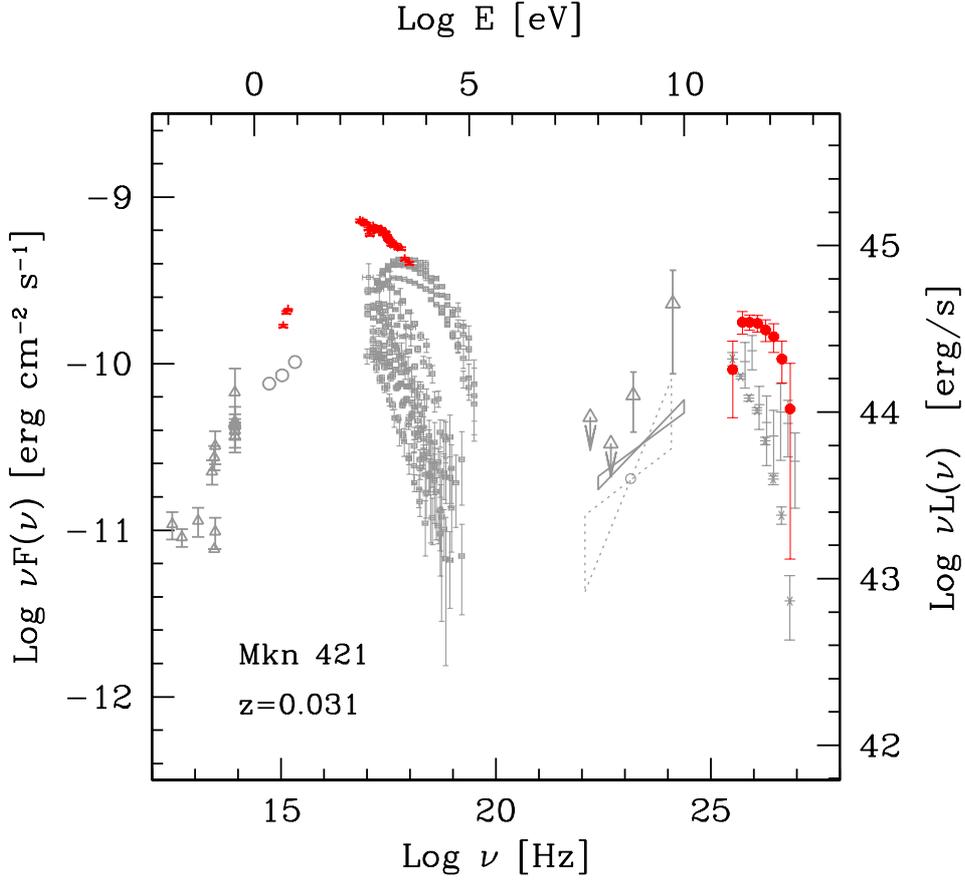} 
  \caption{Example MWL SED of Mrk 421 as observed in tight simultaneity by
    {\it Swift}/UVOT, {\it Swift}/XRT and MAGIC-I on  2008 February 11
    (filled red circles). Historical data  (observed in different campaigns,
      and under less demanding time constraints) taken from
    \citet{2010MNRAS.401.1570T} are plotted for comparison, with gray open
    symbols.}
\label{comparison}
\end{figure*}

Below we focus on the eight cases for which tightly simultaneous observations
in VHE with MAGIC--I telescope and in X--rays with {\it Swift}/XRT could be
performed. 
Table \ref{421_tight_tab} summarizes the observation logs of the
two instruments for these nights. 
For each one we report the beginning and the end of the MAGIC-I
   observation time span, the total effective time, and the ZA range  of each
   dataset. The start time  and duration  of the corresponding  {\it
     Swift}/XRT pointing are also reported, along with the actual overlapped
   observing time (in ks) in the last column.
 The MAGIC--I data considered for each night always cover a timespan that is
 longer than the {\it Swift} exposures: this was
necessary, because the typical observation time of {\it Swift} in this campaign (1 ks) is
enough for deriving a fairly detailed  X--ray spectrum of Mrk 421, but the significantly
lower count rate available in the $\gamma$--ray domain makes this
 exposure time too short for deriving a VHE spectrum detailed enough for the
modeling. Therefore  the whole MAGIC--I exposure was used to
derive the VHE spectrum for each night, given that the observing conditions were stable and no
evidence  for sharp evolution of the source arose from the VHE light curves at
minute scales.  
For each of the eight states we built the MWL SED matching the
MAGIC, {\it Swift}/XRT and optical--UV (either R--band from KVA, or UV
from{\it Swift}/XRT, or both).
As an example, the SED of Mrk 421 as observed on 2008 February 11 is
plotted in Figure \ref{comparison},  compared to  historical MWL data taken from
\citet{2008MNRAS.385L..98T}. It is worth noticing that the VHE SED is
high and hard \citep[in agreement with what is expected from other observations
  of this source during active phases, see e.g.][]{Acciari11}, while the X--ray SED is high but quite soft compared to past
states where the synchrotron peak was observed at higher energies. The wide
separation of the two peaks is discussed in detail in Section \ref{sedmodeling}.
\begin{table*}[th!]
\centering
\begin{tabular}{l|cccc|cr|r}
\hline
\hline
Night \T  \B
&\multicolumn{4}{c|}{MAGIC Obs.}
&\multicolumn{2}{c|}{{\it Swift}/XRT Obs.}
&Overlap\\
\hline
&Start (UT)\T &End (UT)&Time &ZA range& Start (UT)&Time
& Time\\
yyyy-mm-dd&(hh.mm) \B &(hh.mm)&(ks)&(deg)&(hh.mm)&(${\rm ks}$)\phantom{0}&(${\rm ks}$)\phantom{0}\\ 
\hline
 2008-01-08 \T &01.58&02.44&2.7&          31-41&02.30&2.0&0.8 \\
 2008-01-09&03.56&06.19&8.2&\phantom{0}6-20&04.04&2.0&2.0\\
 2008-01-10&02.23&06.05&3.5&          10-34&02.27&2.3&1.1\\
 2008-01-16&03.17&05.11&6.5&\phantom{0}6-21&03.21&1.2&1.2\\
 2008-01-17&03.26&04.25&3.4&\phantom{0}6-18&03.29&0.8 &0.8\\
 2008-02-11&03:33&03:58&1.4&          10-18&03.40&1.9&1.1\\
 2008-04-02&00.45&01.00&0.9&          17-21&00.42&0.9 &0.6 \\
 2008-04-03 \B &21.55&22.20&1.3&          15-23&21.59&1.2&1.1\\
\hline
\hline
\end{tabular}
\caption{Summary of the eight tightly simultaneous observations of Mrk 421 with 
  MAGIC-I and {\it Swift}/XRT.}
\label{421_tight_tab}
\end{table*}

\subsection{SED modeling}
\label{sedmodeling}

To reduce the degrees of freedom, we used a simple one-zone SSC model 
\citep[for details see][]{1998ApJ...509..608T,2003ApJ...593..667M}, similar to the models commonly adopted to reproduce the SED of Mkn 421 \citep[e.g.][]{2004ApJ...601..151K,2008ApJ...686..181F}. 
The emission zone is supposed to be spherical with radius $R$, in motion with bulk Lorentz factor
$\Gamma$ at an angle $\theta $ with respect to the line of sight. Special relativistic effects are 
described by the relativistic Doppler factor, $\delta=[\Gamma(1-\beta \cos \theta)]^{-1}$.
The energy distribution of the relativistic emitting electrons is described
 by a smoothed
broken power law function,  written for better clarity in terms of the
adimensional Lorentz parameter $\gamma=E/m_ec^2$; the distribution spans the
$[\gamma _{\rm min} ,\gamma _{\rm max}]$ energy range, with  slopes $n_1$ and $n_2$ below
and above the break  energy $\gamma _{\rm b}$, respectively. This  purely phenomenological distribution,
with $n_1 < 3$ and $n_2 > 3$, is able to reproduce the observed bumpy SED. 
To calculate the SSC emission we used the full  Klein--Nishina cross section
 \citep{1968PhRv..167.1159J}. 
\begin{table*}[th!]
\centering

\begin{tabular}{lcccccccccc|ccc|cc}
\hline
\hline
Night  \T     & $\gamma _{\rm min}$ & $\gamma _{\rm b}$& $\gamma _{\rm
max}$& $n_1$&$n_2$ &$B$ &$K$ &$R$ & $\delta $ &$t_{var}$&$L_{kin}^{p}$&$L_{kin}^{e}$&$L_{B}$&$u_{e}$&$u_{B}$\\
yyyy-mm-dd \B & $[\cdot 10^3]$ & $[\cdot 10^4]$ &$[\cdot 10^6]$  &  & &$[ G]$ & $[$cm$^{-3}]$  & $[\cdot 10^{15}$cm$]$ &&$[$h$]$&\multicolumn{3}{c|}{$[\cdot 10^{42}$  erg/s$]$}& \multicolumn{2}{c}{$[\cdot10^{-5}$erg/cm$^{3}]$}\\

\hline
2008-01-08 \T &$7.0$&$6.0$&$3.0$&$2.0$&$4.0$&$0.050$&$1700$&$9.0$&$45$&1.8&5.41&91  & 1.61&420 &9.9\\
2008-01-09&$10 $&$2.9$&$3.0$&$2.0$&$4.0$&$0.043$&$3700$&$5.0$&$85$&0.5&7.37&136 & 1.25&600 &7.4\\ 
2008-01-10&$6.0$&$5.7$&$3.0$&$2.0$&$4.0$&$0.037$&$3300$&$5.0$&$70$&0.7&8.83&131 & 0.63&850 &5.4\\
2008-01-16&$8.3$&$6.7$&$3.0$&$2.0$&$4.0$&$0.025$&$4000$&$5.0$&$80$&0.6&9.97&197 & 0.38&980 &2.5\\
2008-01-17&$10 $&$6.0$&$0.7$&$2.0$&$4.2$&$0.037$&$2600$&$7.2$&$60$&1.1&6.18&138 & 0.96&590 &5.4\\
2008-02-11&$11 $&$6.9$&$3.0$&$2.0$&$3.7$&$0.020$&$2400$&$6.6$&$85$&0.7&6.86&187 & 0.47&470 &1.6\\
2008-04-02&$8.0$&$3.2$&$1.0$&$2.0$&$3.5$&$0.050$&$5900$&$3.9$&$70$&0.5&5.24&80  & 0.46&1200 &9.2\\
2008-04-03 \B &$17 $&$20 $&$3.0$&$2.0$&$4.0$&$0.040$&$2000$&$8.5$&$40$&2.0&5.47&120 & 0.62&520 &3.6\\
\hline
\hline
\normalsize
\end{tabular}
\vskip 0.4 true cm
\caption{Input model parameters and derived physical quantities for each of
  the eight simultaneous SED.}
\label{tableparamnotev}
\end{table*}

As emphasized in  \citet{1997MNRAS.292..646B} and \citet{1998ApJ...509..608T},
constraints can be put to this simple model by means of simultaneous multi--wavelength observations.
Indeed, the total number of free parameters of the model is reduced to nine: the six parameters specifying the electron energy distribution plus the Doppler factor, the size of the emission region and the magnetic field. 
On the other hand, from X--ray and VHE observations one can ideally derive seven
observational quantities: the slopes of the synchrotron bump before and above
the peak $\alpha _{1,2}$ (uniquely connected to $n_{1,2}$), the synchrotron
and SSC peak frequencies ($\nu _{\rm s,C}$) and luminosities {$L_{\rm s,C}$}
and the minimum variability timescale $t_{\rm var}$, which provides an upper
limit to the size of the sources through the relation $R<c t_{\rm var}\delta
$. 
 It must be noted that as long as  $\gamma _{\rm min} \ll \gamma _{\rm
  b} \ll \gamma _{\rm  max}$ the  values of $\gamma _{\rm min}$ and $\gamma _{\rm
  max}$ are not very constrained by the observation of the peaks. 
  Nevertheless, the availability of the spectral shape across the instrument
bandpasses and of data at other wavelengths (optical and UV in this case)
provide additional constraints with respect to the simple seven quantities enumerated above.
Therefore, once all the observational quantities are known, one can fairly unambiguosly derive the set of parameters. In this respect, the cases studied
here are quite favorable, because we have a fairly good determination of the peak
frequencies (and fluxes) of both peaks. Indeed, although the synchrotron peak
of Mrk 421 is seldom observed within the band encompassed by XRT, the joint
optical--UV and  X--ray data  provide a good constraint to the position of the
synchrotron peak in all the cases.  The SSC peak is located
either within (see e.g. the SED from February 11 in Figure \ref{comparison}) the
MAGIC-I band, or around its lower edge; in the latter case the pronounced
curvature of the MAGIC--I spectrum at the lowest energies allows us to constrain the peak at energies not much below $\approx 50$ GeV. 

Unfortunately,  for the epochs used to derive the SEDs we do not have
information on the variability timescale, $t_{\rm var}$, one of the key
observational parameters needed to completely close the system and uniquely
derive the parameters. In the X--ray band {\it Swift}/XRT observed in  short
($\sim $ 1 ks) snapshots, while no  convincing evidence for sub--hour variability
arose from the corresponding MAGIC-I observations. Therefore we still have
some freedom in choosing the input parameters: one can obtain different sets
of parameters that reproduce the spectral data equally well but differ in the predicted observed minimum variability timescale.

\begin{figure*}[th]
  \centering
\includegraphics[width=\textwidth]{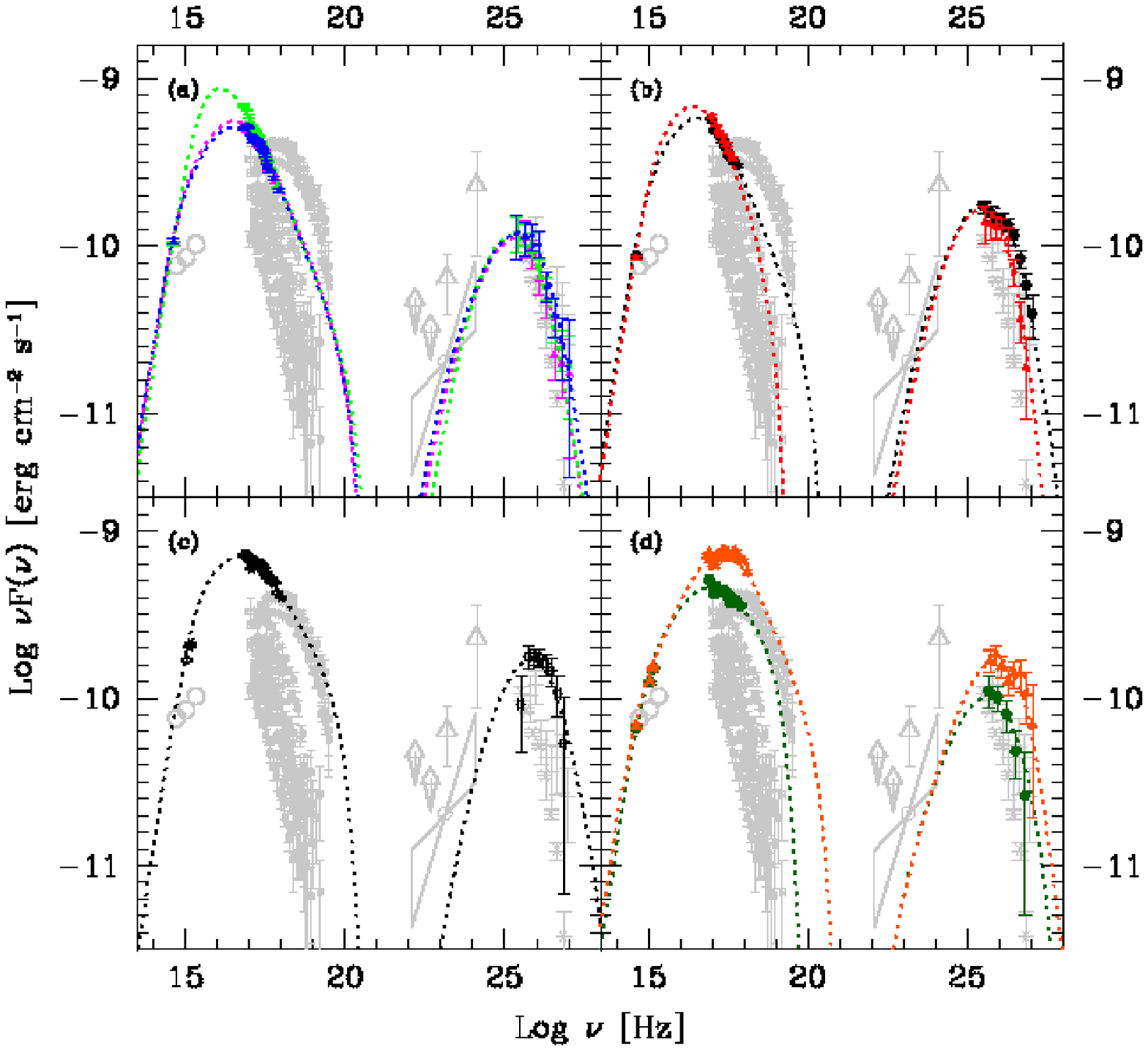} 

  \caption{SED of Mrk 421 with the SSC model overplotted for each of the eight
    simultaneous sets of MAGIC-I, {\it Swift}/XRT and optical--UV data obtained
    in the 2008 campaign. Observation dates are  January 8, 9 and 10 (a),
    January 16 and 17 (b), February 11 (c) and April 2 and 3 (d).}
\label{seds}
\end{figure*}

We applied the model to all eight sets of data collected when {\it Swift} and
MAGIC-I could observe the source simultaneously. 
UVOT and {\it KVA} data were also included in the SED when available. The sets
of parameters obtained from the modeling are reported in Table
\ref{tableparamnotev} and the SED data and the corresponding model are plotted
in Figure \ref{seds}.
 The table reports  for each night
 the minimum ($\gamma_{min}$), break ($\gamma_{b}$) and maximum
 ($\gamma_{max}$) Lorentz factors of the
electron distribution, the low ($n_1$) and high ($n_2$) energy slopes of the electron
distribution, the magnetic field ($B$) and the electron normalization ($K$) within the emitting
region, the radius of the emitting region ($R$) and its Doppler factor ($\delta$). From these
input parameters we derived the  light crossing time ($t_{var}$);  the contributions to
the total jet luminosity  from cold protons ($L_{kin}^{p}$) and
relativistic electrons  ($L_{kin}^{e}$) in the jet,
and from magnetic field  ($L_{B}$); the electron ($u_e$) and magnetic ($u_B$) energy densities.
 We note that in reproducing the SED we did not consider the radio data, since the modeled region
is opaque at these frequencies: in this framework the radio emission
originates in regions of the jet farther away from the black hole, beyond the 
core visible at VLBI scale, which is thought to mark the radio ``photosphere". Accordingly, the inferred source radius is well within the upper limit of $0.1$ pc ($3 \times 10^{17}$ cm) imposed by \cite{2006A&A...457..455C} for the projected size of the SSC zone, based on VLBI observations of the radio core.

Inspection of Table \ref{tableparamnotev} shows that the derived Doppler
factors are quite high, exceeding $\delta=40$ in all the cases and reaching
values as high as 80-85 in the most extreme cases. The main reason for these
high values of $\delta $ is the large separation between the two peaks, the
synchrotron one located below $10^{17}$ Hz, the SSC one around $10^{25}$ Hz or
above. As detailed in, e.g., \citet{2008MNRAS.385L..98T}, a large distance
between the two peaks implies a fairly high value of the Lorentz factor at the peak, $\gamma _{\rm b}$, since $\gamma _{\rm b}=(\nu _{\rm C}/\nu_{\rm s})^{1/2}$, and this directly implies a low $B$ and a large $\delta $ to satisfy the other constraints.

However, we recall that because the variability timescale is not known, we are
left with some freedom in selecting the input parameters. In the models
reported in Figure \ref{seds} we assumed variability timescales in the range 0.5--2 hours,
as typically derived for these sources (see discussion below). In general, the
required Doppler factor roughly scales with the observed variability timescale
as $\delta\propto t_{\rm var}^{-0.5}$
(e.g. \citealt{2008MNRAS.385L..98T}). Therefore, relaxing the condition on
$t_{\rm var}$ and allowing longer
minimum variability timescales, one obtains lower $\delta$.  As an example we used
the case for which we derive the largest $\delta$, that of February 11,
requiring $\delta=85$. As noted above, here the
determination of the peak frequencies is very robust, because the SSC peak falls
 well within the band covered by MAGIC--I. Therefore this is also the best ``benchmark"
available to test the robustness of the derived parameters. For this purpose
we modeled this SED assuming two sets of parameters, basically differing for the
value of the Doppler factor, the radius of the emitting region and the
magnetic field intensity. For  $\delta=85$ we have $t _{\rm var}=0.7$ h
($2.5\times 10^3$ s), while more than halving the Doppler factor $\delta=40$
implies a fairly long variability timescale, $t_{\rm var}=5$ h ($1.8\times
10^4$ s), already longer than the characteristic variability timescale of Mrk
421 in the X--ray band. We can conclude that for the case of February 11,
although the parameters cannot be uniquely fixed, the required Doppler factor
is high, at least higher than $\delta \approx 40$. All other cases are similar.
The derived light crossing times are within the 0.5--2 hours interval. 
This hypothesis matches well the observed typical
raising/decaying timescales of flares of Mrk 421 and similar HBLs (PKS
2155-304, Mrk 501), which are characterized by doubling/halving times of $\approx 10^4$
s (e.g. \citealt{2008ApJ...677..906F}; \citealt{2004A&A...424..841R};
\citealt{2002MNRAS.337..609Z}; \citealt{2000ApJ...543..124T}), with evidence
for the occurrence of even faster events (e.g. \citealt{1996Natur.383..319G};
\citealt{2004ApJ...605..662C}). 

However, if one relaxes this assumption on the variability timescales, the required Doppler factors
remain high. For the data from February 11,  which allow a
quite robust constraint due to direct observation of both the synchrotron and
SSC peak, this implies $\delta > 45$.
  In Table \ref{tableparamnotev} we also report the derived powers carried
  by the different jet
  components, namely cold protons ($L_{kin}^{p}$), relativistic electrons
  ($L_{kin}^{e}$) and  magnetic field ($L_{B}$), assuming a composition of
  one cold proton per relativistic electron. Finally, we computed the electron
  ($u_e$) and magnetic  ($u_B$) energy densities. The jet appears to be strongly
  matter--dominated, as predicted in the standard picture of HBL sources, and in good
  agreement with the result of the modeling performed in \citet{Acciari11}.

\section{Discussion}
\label{discussion}

During the 2008 campaign on Mrk 421 with MAGIC--I a very interesting dataset was
gathered in VHE $\gamma$--rays, complemented by crucial data in optical--UV
and soft X--rays. For the first time it was possible to collect data in these
bands in close simultaneity during high states of the source, so that the
derived spectra sampled the SED close to the synchrotron and IC peaks. In this
situation the parameters describing the source in the framework of the
standard one-zone leptonic model can be determined with remarkable robustness.
One of the most relevant results of our analysis is that to
 reproduce the observed SED with this model, very high Doppler factors are
 required. There is some freedom in choosing the parameters, mainly because of the not known variability timescale at those epochs.
In the models summarized in Table \ref{tableparamnotev} and reported in Figure \ref{seds} we assumed variability
timescales in the range of 0.5--2 hours. 

Indeed, these high values of inferred $\delta $ are not rare: very high
Doppler factors, sometimes higher than $\delta \sim$50, for Mrk 421 and other
well--observed HBLs were obtained in the past, leading to the so called
``$\delta$-crisis" \citep[e.g.][]{2002MNRAS.336..721K,Acciari11,2003ApJ...597..851K,2003ApJ...594L..27G,Henri06,2007A&A...462...29G,2008ApJ...686..181F}.
 Analogously, the recent exceptional VHE flare of PKS 2155-304 \citep{2007ApJ...664L..71A} seems
to require extreme Doppler factors in the framework of one-zone models
(\citealt{2008MNRAS.384L..19B}; \citealt{2008MNRAS.386L..28G};
\citealt{2008ApJ...686..181F}; \citealt{2006ApJ...651..113K}). These high
values of $\delta$ (implying a similarly high value of the bulk Lorentz
factors) contrast with the very low jet velocities inferred at VLBI scales
 in a large part of TeV BL Lacs (e.g. \citealt{2004ApJ...613..752G},
\citealt{2004ApJ...600..115P}), including Mrk 421, and with the value of
$\Gamma$ required from the unification of BL Lacs and FRI radiogalaxies
\citep[e.g.][]{2003ApJ...594L..27G,Henri06}.

In addition to the extreme Doppler factor, we can identify two other problems afflicting the standard model, namely the huge difference between the magnetic and particle energy densities and the extremely long cooling times of the emitting relativistic electrons.

Table \ref{tableparamnotev} shows that in all cases the inferred electron
energy density substantially exceeds the corresponding magnetic energy
density, by up to two orders of magnitude and even more. This result is
 also generally found from SSC fits of HBL SEDs \citep[see e.g.][for a recent example]{Acciari11}, while for other classes of blazars, in particular for FSRQs, equipartition is usually found \citep[e.g.][]{Ghisellini10}. This evidence disagrees with the general expectations of the diffusive shock acceleration models, in which a substantial equipartition between particles and magnetic field is expected. 

Concerning the second problem, namely the long cooling timescales, we point
out that, following for instance the formulae in \citet{1997MNRAS.292..646B}
or \citet{1998ApJ...509..608T}, fairly long cooling times $t_{cool}$,  on the order $10^6$ s in the observer's frame, can be computed from the model parameters in Table \ref{tableparamnotev}. Therefore, the declining part of flares cannot be attributed to the cooling of the emitting electrons. A possibility is that adiabatic expansion, which allows quenching of the flux within scales of $R/c$, has to be invoked as
one of the viable processes that may explain the observed descent of TeV and
X--ray fluxes on hour scales.  But it  must be noted that this explanation
has the significant drawback of implying a  very energetically inefficient
jet. A problem related to the long cooling timescale is that one cannot interpret the observed break in
the underlying electron energy distribution as the separation between fast and
slow cooling particles \citep[e.g.][]{1998ApJ...509..608T}, so it has to be,
for instance, assumed to be intrinsic to the injected population.
These problems could therefore hint at the unsuitability of the one--zone model for this source. 

A solution of these problems faced by the standard one-zone SSC scenario, extensively discussed 
in literature, is based on the possible existence of 
multiple active emitting regions \citep[e.g.][]{Blazejowski05,2003ApJ...594L..27G}.
 based on the possibility that the flow is characterized by
portions moving at different speeds. If these regions emit, in each of them
the electrons can scatter not only the locally-produced synchrotron photons,
but also the soft photons produced in the other region. Moreover, the energy
density of these ``external" photons is amplified in the rest frame of the
emitting region through the relative speed between the two portions of the
flow. The result is that the inverse Compton emission of each region is
amplified with respect to the SSC emission. As a consequence, the Doppler
factor required to reproduce the SED is lower than that of the one-zone
model. In particular, in the ``spine-layer" model of
\citet{2005A&A...432..401G}, it is assumed that the jet has an inner faster
core (the spine) that is surrounded by a slower layer. At a narrow angle of view, which
is characteristic for blazars, the emission is dominated by the faster spine whose IC
emission is a mixture of SSC and ``external" Compton components.  

This model
would also more easily accomodate the short variability time scales observed
in Mrk 421, which are hardly explained within the one--zone model  due to the
long electron cooling times, as discussed in Section \ref{sedmodeling}. Indeed, in
general a 10 times more intense $B$ can be adopted when modeling a given SED (see
e.g. \citealt{2005A&A...432..401G}); because the
 synchrotron cooling time scales as $t_{sync} \propto B^{-2}$, this could lead to cooling
times on the order of the required variability time scale.

An alternative scenario that is possibly able to solve these problems is the ``minijets" model advocated by \citet{2009MNRAS.395L..29G,2010MNRAS.402.1649G}. In this framework, the emission is thought to occur in very fast  small portions of plasma resulting from the rapid reconnection of magnetic field lines inside the main jet  flow. If the magnetization (ratio of magnetic over kinetic jet power) is high enough, the Lorentz factor of these blobs in the rest frame of the jet can be as high as $\Gamma=50$. Moreover, since the emitting plasma is the residual of the annihilation of magnetic field, one naturally expects a low magnetic energy density and thus a high particle over magnetic energy density ratio.

A modeling of the SED with the more complex (and less constrained)  models mentioned above is beyond the scope of this paper and left to future work.

\begin{acknowledgements}

We would like to thank the Instituto de Astrof\'{\i}sica de
Canarias for the excellent working conditions at the
Observatorio del Roque de los Muchachos in La Palma.
The support of the German BMBF and MPG, the Italian INFN, 
the Swiss National Fund SNF, and the Spanish MICINN is 
gratefully acknowledged. This work was also supported by 
the Marie Curie program, by the CPAN CSD2007-00042 and MultiDark
CSD2009-00064 projects of the Spanish Consolider-Ingenio 2010
programme, by grant DO02-353 of the Bulgarian NSF, by grant 127740 of 
the Academy of Finland, by the YIP of the Helmholtz Gemeinschaft, 
by the DFG Cluster of Excellence ``Origin and Structure of the 
Universe'', and by the Polish MNiSzW grant 745/N-HESS-MAGIC/2010/0.
We acknowledge insightful and constructive criticism from the anonymous
referee.
\end{acknowledgements}

\end{document}